\documentclass[twocolumn, twocolappendix]{aastex631}

\usepackage{amsmath}
\usepackage{amssymb}
\usepackage{bm}
\usepackage{xcolor}
\usepackage{enumitem}

\received{March 4, 2025}
\revised{January 4, 2026}

\newcommand{\parens}[1]{\left(#1\right)}
\newcommand{\brackets}[1]{\left[#1\right]}
\newcommand{\chevronsi}[1]{\langle#1\rangle}
 
\newcommand{\unit}[1]{\hat{\bm{#1}}}

\graphicspath{{./}{figs/}}
\begin{document}

\title{A Physical Model of Pulsar X-ray Filaments}

\author[0000-0002-6401-778X]{Jack T. Dinsmore}
\affiliation{Department of Physics, Stanford University, Stanford CA 94305}
\affiliation{Kavli Institute for Particle Astrophysics and Cosmology, Stanford University, Stanford CA 94305} 
\author[0000-0001-6711-3286]{Roger W. Romani}
\affiliation{Department of Physics, Stanford University, Stanford CA 94305}
\affiliation{Kavli Institute for Particle Astrophysics and Cosmology, Stanford University, Stanford CA 94305}

\begin{abstract}
We present a model for pulsar filaments---a class of narrow X-ray nebulae misaligned with the proper motion, powered by pulsar-generated $e^\pm$. We suggest that cosmic ray-enhanced turbulence drives pitch-angle scattering and dominates $e^\pm$ motion along the filament; highly amplified magnetic fields are not required. A simulation built on this picture, using analytic approximations for the turbulence growth and cosmic ray evolution, generates images and spectra matching observations of the three best-measured filaments. The model structure depends on interstellar medium properties, and fits to filament data require values similar to observed ISM values. In this model a substantial fraction of the filament $e^\pm$ escape, free-streaming for many pc, in contrast to the suppressed cosmic ray diffusion near pulsar TeV halos. Accordingly, nearby low-power filament-generating pulsars may make out-sized contributions to the local positron spectrum. Future X-ray observatories can make the sensitive spectral maps required to test this particle escape.
\end{abstract}


\section{Introduction} \label{sec:intro}
X-ray observations of pulsars moving at supersonic speeds through the interstellar medium (ISM) show several examples of narrow, extended X-ray structures oblique to the proper motion. The first ``pulsar X-ray filament'' or ``misaligned outflow'' was discovered emanating from PSR B2224+65 \citep[with the ``Guitar'' H$\alpha$ nebula,][]{cordes1993guitar,wang2021xray,de2022quarter}; the brightest extends from PSR J1101$-$6101 \citep[Lighthouse,][]{pavan2014long,tomsick2012is,klingler2023nustar}; and the narrowest from PSR J2030+4415 \citep{de2020psr,de2022long}. PSR J1509$-$5850 \citep{hui2007radio,klingler2016chandra}, PSR J2055+2539 \citep{marelli2016tale,marelli2019two}, and PSR J1957+5033 \citep{dinsmore2026chandra} also display clear X-ray filaments, and more candidates are discussed in \cite{dinsmore2024catalog}.

The X-ray emission is likely synchrotron radiation from ultrarelativistic electrons and positrons ejected onto ISM magnetic field lines, as first suggested by \cite{bandiera2008on}. Measurements in \cite{churazov2024pulsar} and \cite{dinsmore2025starlight} support this hypothesis, finding the local magnetic field to be aligned with the X-ray structure in a filament candidate and a filament, respectively. The acceleration and escape of these leptons from the pulsar bow shock have been modeled by several authors \citep[e.g.][]{barkov20193d}. The post-escape propagation within the filament was considered by \citet{2017SSRv..207..235B} who suggested that injected cosmic rays (CRs) generate turbulence via the non-resonant streaming instability \citep[NRSI;][]{bell2004turbulent}, which creates turbulence at smaller length scales than those responsible for particle scattering. This picture has recently been developed in an analytic model by \cite{olmi2024nature}. In this model, CRs flow unperturbed until the instability saturates, and this saturation time sets the filament length. After saturation, the field is highly turbulent with strength elevated to $26 -181$ $\mu$G, well above initial values. Although attractive, this model has some challenges. In particular, the CRs propagate at almost the speed of light $c$, whereas magnetic turbulence should propagate near the Alfv\'en velocity $v_A \ll c$. The turbulence at the end of the filament is therefore isolated from the turbulence near the base; they have different amplitudes and saturate at different times. This means that a single-zone picture of the filament is inadequate, challenging the notion of a global saturation time setting the filament length. Furthermore, equipartition estimates for the Guitar filament do not support such high $B$ values \citep{de2022quarter}. Similarly, the filament candidate G0.13$-$0.11 exhibits high X-ray polarization degree, implying only moderate magnetic turbulence in the radiation zone \citep{churazov2024pulsar}. Thus, it would be attractive to find an alternative model that considers spatially dependent turbulence growth and exhibits low-amplitude turbulence in steady state.

We propose that pitch angle scattering from the resonant streaming instability (RSI) causes filament CRs to slow and reflect. The reflection rate is initially low, but grows exponentially because reflected CRs dilute the strong pulsar-injected current, suppressing the NRSI and enhancing the RSI. The exponential turbulence growth is cut off at still-linear intensities because particle injection into a given field line ceases when the pulsar advances to fresh, unperturbed ISM fields. Because of the RSI's enhancement, CRs can be scattered without requiring the NRSI to reach saturation. We thus avoid the associated strong magnetic fields and strong turbulence.

\S \ref{sec:argument} describes this model and its assumptions at a qualitative level. The model's quantitative realization is given in \S \ref{sec:model}, including a particle-based numerical simulation, which rapidly computes the filament radiation and provides detailed X-ray filament predictions after a fit to data. In \S \ref{sec:results} we compare the simulation images and surface brightness profiles with the most prominent observed examples, considering both steady and intermittent CR injection. \S\ref{sec:predictions} discusses the general properties of the model, including spectral maps. We conclude in \S \ref{sec:conclusion}, describing how more detailed computations and simulations might test some of our physical assumptions, and how comparison with future filament observations can further test the model.

\section{Summary of the Model}
\label{sec:argument}

The basic assumptions of our model are as follows. We provide justifications at the end of this section. 
\begin{enumerate}[label=(\Roman*)]
  \item CRs are injected into the filament from the bow shock apex. The injection spectrum is power law, low pitch angle, and charge-separated. The rate is chosen such that the simulated steady-state filament luminosities match the observed values. 
  \item CRs stream along magnetic field lines and pitch-angle scatter off resonant turbulence.
  \item CRs also {\it generate} (anisotropic) turbulence, via the NRSI in strong currents near the pulsar and via the RSI in regions of low current density.
  \item This turbulence is stationary in the ISM rest frame and perturbative. It varies spatially due to the spatial variation of CR/current density.
  \item Standard synchrotron radiation applies, including beaming and particle cooling.
  \item Particle-injected turbulence inefficiently cascades to larger scales. This allows CRs to slowly diffuse between field lines.
\end{enumerate}
These principles guide our simulation design (\S\ref{sec:model}), which generates predictions that match observations, such as the filament length and width (\S\ref{sec:results}). The qualitative process by which filaments grow is as follows. We list the assumptions critical to each step in parentheses.
\begin{enumerate}[label=(\Alph*)]
  \item \textit{Initial epoch}: Injected CRs free-stream from the pulsar through the ISM (I). Reflection occurs (II), but is initially quite limited, since the standard $\gtrsim$ 10 pc ISM mean free path is much longer than the filament. The injection zone has large current density, so the NRSI generates turbulence along the filament leading edge at length scales decoupled from the resonant scale (III).
  \item \textit{Turbulence growth}: Reflecting particles return toward the injection zone, lowering the current density starting at the far end of the leading edge (II, VI). As the RSI takes over, the turbulence injection scale approaches the resonant scale, and low energy CRs resonantly scatter off high energy CR NRSI turbulence (III). This boosts the returning fraction; the feedback exponentially grows the scattering rate at the leading edge, especially away from the pulsar injection site.
  \item \textit{Growth cutoff}: As the pulsar shifts to unperturbed magnetic field lines, particles are no longer injected into the strip of turbulence generated in steps A and B. The turbulence growth rate slows dramatically and is no longer exponential. Traversal to a new ISM field line, where injection continues, is faster than the NRSI saturation time.
  \item \textit{Steady state turbulence distribution}: After an initial transient, steady state is reached with the NRSI acting close to the pulsar and RSI taking over farther away and behind the leading edge. This scattering creates a long, narrow strip of turbulence, which builds from low native ISM levels ahead of the pulsar to higher (but still linear) levels in the bulk of the filament behind the leading edge. The rate of turbulence growth (step B) and the cutoff time (step C) set the filament length.
  \item \textit{Steady state particle distribution}: A substantial fraction of the particles do not reflect within the filament and escape, propagating to the unperturbed ISM (II, IV), producing little X-ray synchrotron emission in a low surface brightness zone. The reflected CRs scatters in the turbulence behind the leading edge, radiating until they either random walk into the low-turbulence zone at the end farthest from the pulsar and escape (primarily along the background field via II), or cool out of the \textit{Chandra} band (V). The rates of these processes set the filament width.
\end{enumerate}

Our justification for each assumption follows. Possible model-affecting weaknesses are discussed in \S\ref{sec:parameters}. 
(I) The injection spectrum slope is chosen to match the observed X-ray spectrum, which is inconsistent with monoenergetic injection.
As in \cite{olmi2024nature}, we inject CRs with low initial pitch angle $\alpha$ because the conserved adiabatic invariant should reduce $\alpha$ as CRs flow away from the highly magnetized bow shock. Charge separation is motivated by the simulations of \cite{olmi2019full}, and was assumed by \cite{olmi2024nature}. \S\ref{sec:turbulence} further discusses this assumption. The injection rates we need to match observed filament luminosities in \S\ref{sec:results} are a few $\times 10^{-4} \dot E$, where $\dot E$ is the pulsar spin-down luminosity. The injection rate is therefore energetically acceptable.
(II) Modeling the pitch angle population is important because the CR motion, scattering rates, and the turbulence injection rate/scale all depend strongly on pitch angle. An analytic expression for this scattering in perturbative turbulence is available, allowing us to follow the pitch angle distribution's evolution in the linear regime (\S\ref{sec:particle}).
(III) As \cite{olmi2019full} point out, the pulsar outflow creates strong current density near the injection site if charge separation occurs. Here turbulence is generated via the NRSI. However, CR reflection (via scattering) means that most of the filament volume has low current density, so the RSI dominates. We use a momentum conservation argument, like that employed by \cite{kulsrud2004plasma} for the RSI, to determine turbulence growth rates as a function of the local current density. Thus we can self-consistently model the RSI and NRSI (\S\ref{sec:turbulence}) and their spatial variation, taking pitch angle into account.
(IV) \cite{kobzar2017spatio} have conducted particle-in-cell simulations of NRSI growth from constant-current CR injection in a non-periodic box, and found that turbulence first saturates in the region that first receives the driving current. This is particularly relevant to filaments, as the NRSI would saturate first near the pulsar and could halt flow to the end of the filament, destroying the filamentary structure. It is therefore important to model spatially dependent turbulence. Since \cite{churazov2024pulsar} saw low-amplitude turbulence in G0.13$-$0.11 and turbulence propagates at $v_A$ (much slower than the CRs or the pulsar velocity), our assumptions of perturbative and stationary turbulence are also justified.
(V) The synchrotron treatment is standard. Since the radiation is beamed at $\alpha$ to the field line, the filament brightness is viewing angle dependent. Synchrotron cooling (\S\ref{sec:particle}) is significant for the CRs retained in the filament.
(VI) Although turbulence is perturbative, small, non-linear effects should lead to slow cascades to both larger and smaller scales \citep{schroer2025role}. Such effects are not explicit in our linearized simulation, so we introduce a parameter to describe the resulting cross field diffusion, which we fit to characterize its efficiency (\S\ref{sec:turbulence}).

Our simulation is intended to be a simple tool, efficient enough that we can explore the effects of the model parameters and fit these to data. Ultimately a full simulation that properly resolves all scales and non-linear effects will be necessary to validate these assumptions, though this will present numerical challenges.

\section{Model details}
\label{sec:model}
To model the detailed turbulence structure of the leading edge, we must resolve scales from the bow shock stand-off distance (hundreds of AU) to the filament length ($\sim$ parsec). The CR distribution function must be five-dimensional, modeling location, energy and pitch angle to properly treat assumption II. Assumption III requires the turbulence distribution function to be four-dimensional, depending on position and wavenumber. Furthermore the sharp structure of the filament leading edge, which is critical to our model, should not be erased by numerical diffusion.

To accomplish this, we simulate the CRs as particles. Scattering is simulated by random walking their pitch angle and position using the analytically computed diffusion constants presented below. Turbulence modeling is simpler as it is assumed to be stationary in the ISM frame; we evolve it as a fluid in a grid.

\subsection{Particle Model}
\label{sec:particle}

We write the magnetic field as $\bm B(\bm x) = B_0(\unit z + \bm b(\bm x)$), with unitless turbulent magnetic field $\bm b(\bm x)$ which randomly varies as a function of position. Per assumption IV, $|\bm b| \ll 1$. We also assume a ``slab'' field geometry with $\bm b$ transverse to the background field $B_0 \unit z$, and wavenumber $\bm k$ parallel to $\unit z$. The spectral energy density $s_\tau(k)$ of the unitless $\bm b$ field summarizes all the first-order properties of turbulence (see Eq.~\ref{eqn:tensor} for the definition). In particular, the fraction of magnetic energy density due to turbulence is
\begin{equation}
  u_\tau \equiv \frac{B^2 / 8\pi}{B_0^2 / 8\pi} -1 = \chevronsi{b^2}= \int_0^\infty dk\, s_\tau(k).
  \label{eqn:u_tau}
\end{equation}

\paragraph{Particle motion} The diffusion constants, pitch angle $\alpha$ and perpendicular motion of a CR moving in this field have been worked out using quasilinear theory \citep{jokipii1966cosmic,earl1974charged, matthaeus2003nonlinear,shalchi2005second} and validated for high energy particles in simulations \citep{giacalone1999transport,mace2000numerical,shalchi2005second, shalchi2009analytical}. These results with our magnetic field geometry assumptions gives diffusion coefficients for $\cos \alpha$ and the position perpendicular to the background field
\begin{align}
  D_{\cos \alpha} &= \frac{\pi \sin^2 \alpha}{2} k_\mathrm{res}s_\tau(k_\mathrm{res}) \omega_B  \label{eqn:d-cos-alpha} \\
  D_x=D_y &= \frac{\pi \cos^2 \alpha}{2} k_\mathrm{res}s_\tau(0) a_B^2 \omega_B  \label{eqn:d-perp}
\end{align}
where $k_\mathrm{res} = \omega_B/(c|\cos \alpha|)$ is the wavenumber resonant with the CR Larmor radius, $\omega_B$ is the orbit frequency, and $a_B = c / \omega_B$ is the Larmor radius of an $\alpha=90^\circ$ particle.\footnote{These follow from \cite{jokipii1966cosmic} Eq.~37, 38 with $P_{ZZ}=0$ for our slab geometry.}
The drift coefficients associated with parameter $x$ are $\mu_x = \partial D_x / \partial x$ \citep{jokipii1966cosmic}.

A well-studied formal problem of this first-order approach is that, as $\alpha$ approaches 90$^\circ$, $D_{\cos \alpha}$ approaches zero because $k_\mathrm{res}$ approaches infinity. In practice, second order effects allow CRs to reflect regardless. We use the following approach to solve this problem in our model: turbulence changes the local magnetic field direction so that the ``effective pitch angle'' $\alpha'$ between the \textit{local} field and the CR velocity is not the same as the angle between the \textit{background} field and the CR velocity $\alpha$. We assume that $\alpha'$ is what sets the scattering behavior. The angle of the local magnetic field is determined by the power spectrum of transverse magnetic fluctuations $s_\tau(k)$. Averaging over space gives a variance of $\cos \alpha'$ equal to
\begin{equation}
    \chevronsi{(\cos \alpha' - \cos \alpha)^2} = \frac{1}{2} u_\tau\sin^2 \alpha,
  \label{eqn:effective-pitch-angle}
\end{equation}
which is largest for large $\alpha$ CRs in turbulent regions. This approach to resolving the mirror reflection problem gives similar results to the second order approach found in \cite{shalchi2009analytical}. We use Eq.~\ref{eqn:effective-pitch-angle} for the sake of consistency, since the same technique can be used to predict synchrotron beaming in turbulent fields.

\paragraph{Radiation}
\label{sec:radiation}
Particles synchrotron radiate with the typical power for the background field strength $B_0$, plus a small correction due to turbulence derived in appendix \ref{app:b}. Adding these effects, the radiated power is
\begin{equation}
    L=\frac{2e^2}{3m^2 c^3} B_0^2 \gamma^2 \beta^2 \sin^2 \alpha \brackets{1 + u_\tau \parens{1 - \frac{\sin^2 \alpha}{2}}}.
    \label{eqn:sync}
\end{equation}
We use Eq.~\ref{eqn:sync} to simulate particle cooling by reducing the particle energy as $\dot E = -L$.

To calculate the observed synchrotron luminosity, we first compute the spectrum
\begin{equation}
    \frac{dP}{dE_\gamma}(E_\gamma) = \frac{\sqrt{3}}{2\pi} \frac{e^3 B_0 \sin \alpha}{mc^2} \frac{E_\gamma}{\hbar\omega_c}\int_{E_\gamma/(\hbar\omega_c)}^\infty d\xi\,  K_{5/3}(\xi)
    \label{eqn:sync-spec}
\end{equation}
\citep{rybicki2024radiative} where $\omega_c = 3\omega_B \gamma^2 \sin \alpha / 2$, and $K_{5/3}$ is a modified Bessel function of the second kind. No turbulent amplification is included in Eq.~\ref{eqn:sync-spec} because turbulence-induced synchrotron radiation is generally deposited at higher frequencies than the radiation from the background field detected by \textit{Chandra} \citep{toptygin1987role}. In any event the added power from the turbulent field is small in our simulated filaments. Synchrotron radiation is beamed into the surface of a cone with opening half-angle $\alpha'$ from Eq.~\ref{eqn:effective-pitch-angle}, and the beam is blurred by the probability distribution of $\alpha'$. This amplifies the flux as viewed from inclination angle $\iota$ by
\begin{equation}
  \eta = \frac{1}{\sqrt{\pi u_\tau \sin^2 \alpha}} \exp\brackets{-\frac{\parens{\cos \iota - \cos \alpha}^2}{u_\tau \sin^2 \alpha}},
  \label{eqn:beaming}
\end{equation}
suppressing visibility of low-$\alpha$ CRs. To compute the observed luminosity, we integrate Eq.~\ref{eqn:sync-spec} over the 0.5--7 keV \textit{Chandra} band and multiply by the beaming factor $\eta$.

\subsection{Turbulence Model}
\label{sec:turbulence}

NRSI and RSI growth rates are usually calculated through the relativistic Vlasov equation, but because we use a particle simulation we cannot compute the growth rate in this way. Fortunately, \citep{kulsrud2004plasma} showed that the RSI turbulence injection rate can be calculated to a good approximation by ascribing the momentum shed by scattering CRs to newly generated Alfv\'en waves. \S\ref{sec:particle} derived the momentum shed by CRs, so we can employ this picture to self-consistently calculate the turbulence growth rate.

\paragraph{Turbulence growth rate}
We assume each CR injects turbulence into a single wavenumber $k_\mathrm{inj}$, calculated later. Consider a turbulence grid cell with volume $\Delta V$ containing the turbulent energy in a range $\Delta k_\mathrm{inj}$. Suppose there are $N$ CRs within this cell with $k_\mathrm{inj}$ in the range. The longitudinal CR momentum is $p_z = NE\cos \alpha/c$ where $E$ is the CR energy. $\cos \alpha$ evolves diffusively with diffusion constant $D_{\cos \alpha}$ (Eq.~\ref{eqn:d-cos-alpha}). Changing variables to $p_z$ shows that the CR momentum evolves with diffusion constant $D_{p_z} = (NE/c)^2 D_{\cos \alpha}$.  Meanwhile, the Alfv\'en wave momentum $p_A$ is related to the wave energy through $E_A = p_Av_A$, which is related to the turbulent energy density by $E_A = U_{B_0}s_\tau(k_\mathrm{inj}) \Delta V \Delta k_\mathrm{inj}$ from the definition of $s_\tau$ (Eq.~\ref{eqn:tensor}). When the CR momentum is transferred to turbulence, the random walk of $p_z$ will lead to a random walk in $p_A$ and therefore $s_\tau$. By a change in variables, the diffusion constant for $s_\tau$ is

\begin{equation}
  D_{s_\tau} = \parens{\frac{NE \beta_A}{\Delta V \Delta k_\mathrm{inj} U_{B_0}}}^2 D_{\cos \alpha}
\end{equation}
where $\beta_A = v_A /c$. Only the long-term growth of $s_\tau = \sqrt{2 D_{s_\tau} t}$ matters for this simulation. Differentiating this expression allows us to compute the $s_\tau$ growth rate in terms of quantities accessible in our simulation:

\begin{equation}
  \dot s_\tau(k_\mathrm{inj}) = \frac{1}{s_\tau(k_\mathrm{inj})}\parens{\frac{n_k\beta_A}{U_{B_0}}}^2 \chevronsi{E^2D_{\cos \alpha}}.
  \label{eqn:e-tau-dot}
\end{equation}
where $\chevronsi{-}$ represents an average over particles in the grid cell and the CR phase space number density is $n_k = N/(\Delta V \Delta k_\mathrm{inj})$.

\paragraph{Injection wavenumber} In the RSI, $k_\mathrm{inj} = k_\mathrm{res}$. In the presence of a strong CR current, the NRSI applies and $k_\mathrm{inj} > k_\mathrm{res}$. \cite{bell2004turbulent} arrives at a range of $k_\mathrm{inj}$ by deriving the modification to the MHD equations of motion and applying the Vlasov equation. In a particle-based simulation where pitch angle is important, we cannot use the same procedure. To approximate the injection scale, we calculate how the current's contribution to Alfv\'en wave dispersion relation shifts $k_\mathrm{res}$ at fixed frequency. We inject turbulence into the shifted wavenumber $k_\mathrm{inj}$.
\cite{bell2004turbulent} derives an Alfv\'en wave dispersion relation $\omega^2 - k^2v_A^2 = \pm 2\xi v_A^2 k / a_B$, where $\xi=2\pi j_z / (\omega_B B_0)$ for current density $j_z$.\footnote{\cite{bell2004turbulent}, Eq.~15. Our $\xi$ is Bell's $\zeta v_s^2/(2v_A^2)$.} In quantities followed in our simulation,
\begin{equation}
  \xi = \frac{1}{4} \frac{U_K}{U_{B_0}} \chevronsi{\cos \alpha}
  \label{eqn:xi}
\end{equation}
where $U_K$ is the CR kinetic energy density. Eq.~\ref{eqn:xi} assumes total charge separation at injection. If charges are allowed to mix, $j_z$ and therefore $\xi$ is reduced by an additional charge mixing factor. As long as $\xi$ remains greater than a few in the injection zone, filament morphologies are realistic. Our simulation therefore requires some charge mixing. But if charges are completely mixed and so $j_z=\xi=0$ (i.e.~the RSI is dominant everywhere), then particles scatter off self-generated turbulence before propagating significantly at the leading edge, and filament collapses into a compact PWN.

Choosing a frequency that reproduces the RSI for no current, the wavenumber receiving turbulence is
\begin{equation}
  k_\mathrm{inj} = \frac{1}{a_B} \parens{\xi + \sqrt{\sec^2 \alpha + \xi^2}}.
  \label{eqn:k-inj}
\end{equation}
For large current density, $k_\mathrm{inj} = 2\xi/a_B$. \cite{bell2004turbulent} reports $k_\mathrm{max} = \xi/r_L$, which is equivalent for typical pitch angles with $\cos \alpha = 1/2$. Eq.\,\ref{eqn:k-inj} therefore describes the RSI and NRSI turbulence scales in a single formula with the correct limits at $\zeta \rightarrow 0, \infty$.

\paragraph{Alfv\'en velocity value} The Alfv\'en velocity in units of $c$ is $\beta_A=B_0/\sqrt{4\pi \mu m_p n}$, where $\mu\approx 1.3$ is the average ion molecular mass, $m_p$ is the proton mass, and $n$ is the number density of particles coupled to the wave. Only ions couple directly to the wave through the Lorentz force, but neutrals can also couple indirectly if the collision rate $\nu$ is greater than the Alfv\'en wave frequency $\omega_A$ \citep{drury1996limits,nava2016nonlinear}. For the $\gamma \sim 10^7$ particles responsible for most of the turbulence, ambipolar diffusion is small and neutrals are effectively decoupled: $\omega_A / \nu = 1.7 B_{10}^2 \gamma_7^{-1} n_{0.5}^{-3/2} f_{5}^{-1/2} T_4^{-0.4}$. Here we use the damping rate $\nu$ of \cite{kulsrud1971effectiveness} \citep[see][]{drury1996limits}, total ISM density $0.5 n_{0.5}\ {\rm cm ^{-3}}$, warm neutral medium (WNM) ionization fraction 0.05$f_5$ \citep[e.g.][]{wolfire1995neutral, jenkins2013fractional}, and WNM temperature $\lesssim10^4 T_4$ K. Thus the ion number density is used here to approximately compute Alfv\'en velocity.

\paragraph{Other considerations} We use Kraichnan scaling $s_\tau(k) \propto k^{-3/2}$ for the ISM (initial) turbulence spectrum because it gives $D_{\cos\alpha} \propto E^{-1/2}$ scaling with the CR energy $E$ (Eq.~\ref{eqn:d-cos-alpha}), and therefore $D_\parallel \sim 1/D_{\cos\alpha} \propto E^{1/2}$ for the ISM CR diffusion coefficient $D_\parallel$. This is consistent with observations (see \S\ref{sec:parameters}),
though other spectral indices are also theoretically motivated. 
We also tested a Kolmogorov $k^{-5/3}$ spectrum and saw only minor differences to the model output. The initial amplitude $u_\tau=2.2 \times 10^{-3}$ is computed using Eq.~\ref{eqn:d-cos-alpha} from our choice for $D_\parallel$, which we discuss in \S\ref{sec:results}.

To linear order, this injected turbulence cannot affect cross-field diffusion ($D_x$ and $D_y$) since these are determined by larger scale fluctuations. However, nonlinear effects will initiate a turbulent inverse cascade, redistributing a small fraction of the injected turbulence energy to these large scales \citep{schroer2025role}. 
We cannot model such non-linear effects in detail with the present linear model, so we describe the cascade strength with a free parameter
\begin{equation}
    f_\perp \equiv s_\tau(0) / s_\tau(k_\mathrm{res,min})
    \label{eqn:f-perp}
\end{equation}
where $k_\mathrm{res,min}$ is the smallest $k_\mathrm{res}$ of the injected particle population. One expects $f_\perp \ll 1$.

\subsection{Particle Injection}
\label{sec:injection}

We inject particles into a tube of radius 100 AU, comparable to the shock stand-off distance $r_0 = [\dot E/(4\pi c \rho_\mathrm{ISM} v_\mathrm{psr}^2)]^{1/2}$, where $v_\mathrm{psr}$ is the pulsar velocity. Our particle spectrum $dN/dE = E^{-2}$ emitting in the $0.5-10$ keV range covers the \textit{Chandra} band.\footnote{Observational constraints on the injected energy range are weak at present. The most energetic filament photon detected is $\sim 25$ keV \citep{klingler2023nustar}. The lack of radio detection argues for a sharp low energy cut-off as well. The variability of the Guitar filament suggests that $r_0$ fluctuations dramatically affect the injected power, which should occur if the injected energy range is small, e.g.~a decade or less \citep{de2022long}.} The injection rate is set to reproduce the observed filament luminosity, and is always a small fraction of the pulsar spin-down power.

As mentioned in \S \ref{sec:argument}, the field strength transitions from the strong $(4\pi \rho_{\rm ISM})^{1/2}v_{\rm psr}\sim$mG accelerating shock fields to the relatively weak $B_0 \sim 10\, \mu$G ISM field along the CR trajectory. If the transition is adiabatic, $\sin^2 \alpha / B$ must remain constant so that the maximum possible pitch angle is $\sim \sqrt{\epsilon_B}$, where $\epsilon_B = B_0 / B_\mathrm{shock} \ll 1$. Thus  evolution from an isotropic pre-injection pitch angle distribution gives.
\begin{equation}
  P(\cos \alpha) = \begin{cases}
      \brackets{\frac{1}{\epsilon_B} \frac{\cos^2 \alpha}{\epsilon_B - \sin^2 \alpha}}^{1/2} &  \alpha < \sin^{-1} \sqrt{\epsilon_B}\\
      0 & \mathrm{otherwise}\\
  \end{cases}.
\end{equation}
We take $\epsilon_B = 5\times 10^{-2}$.

Some initial pitch angle anisotropy is critical to this model. If particles are injected with isotropic pitch angle ($\epsilon_B = 1$), most particles generate turbulence near the injection point resulting in compact emission near the bow shock apex, rather than the long, near-uniform brightness profile of observed filaments. However, as long as $\epsilon_B \lesssim 0.2$, we find the value only weakly affects filament morphology.

Once particles have scattered within the filament and the pitch angle distribution returns to isotropy, by the same logic they cannot return to the bow shock or the PWN trail. We therefore assume a mirror-reflecting boundary on the PWN side of the filament. Other CR boundary conditions are open.

\subsection{Simulation design}
\label{sec:simulation}

We implement a three-dimensional filament simulation following the processes described above. After injection, the particles propagate along $\unit z$ (the background field direction) at speed $c\cos \alpha$. To model diffusion of a CR parameter $x$ (e.g.~$\cos \alpha$), we update $x$ after a small time step $\Delta t$ by $\mu_x\Delta t + g \sqrt{2 D_x\Delta t}$, where $g$ is a random variable drawn from the standard Gaussian distribution and $\mu_x, D_x$ are the drift and diffusion coefficients. These are given by Eqs.~\ref{eqn:d-cos-alpha}, \ref{eqn:d-perp}, and \ref{eqn:f-perp} and evaluated with pitch angle $\alpha'$ such that $\cos \alpha'$ is drawn from a Gaussian distribution with variance given in Eq.~\ref{eqn:effective-pitch-angle}.

We bin the turbulent spectrum $s_\tau$ into a three-dimensional spatial grid. Near the leading edge, we set the grid bins' cross-field  half-width to the Larmor $r_L$ radius of the highest energy injected CRs. This is to achieve the highest resolution possible in this important region without modeling the CRs' helical motion (radius $r_L$). In the main body of the filament, turbulence is well developed with small spatial gradients, and high resolution is less critical. There we use larger bins to improve computational efficiency. We lock this non-uniform grid to move with pulsar and advect turbulence away from the leading edge with velocity $-\bm v_\mathrm{psr}$ so it remains stationary with the ISM. This is done by evolving the turbulence density in bin $i$ with width $w_i$ as
\begin{equation}
    (\Delta s_\tau)_i = (s_\tau)_{i-1} \frac{v_\mathrm{psr} \Delta t}{w_{i-1}} -(s_\tau)_i \frac{v_\mathrm{psr} \Delta t}{w_i} + \Delta t (\dot s_\tau)_i.
    \label{eqn:turbulence-propagation}
\end{equation}
The first term receives turbulence from the upstream bin $i-1$, the second term advects turbulence from the current bin downstream, and the last term incorporates the turbulence growth of Eq.~\ref{eqn:e-tau-dot}. Eq.~\ref{eqn:turbulence-propagation} conserves total turbulence energy. It leads to some numerical diffusion of turbulence behind the leading edge; however, turbulence is already quite smooth in this region. By varying the simulation grid size, we determine that numerical diffusion does not affect our results. We also varied other meta-parameters such as $\Delta t$, the number of simulated particles, the number of spatial and wavenumber bins in the turbulence grid, and the locations of the spatial, wavenumber, and energy bounds of the simulation to ensure the results do not depend on them.

Our simulation self-consistently follows feedback between turbulence and the particles.  The rest of this work draws conclusions from X-ray images generated using this simulation.

\section{Comparison to Data}
\begin{figure}
  \centering
  \includegraphics[width=0.95\linewidth]{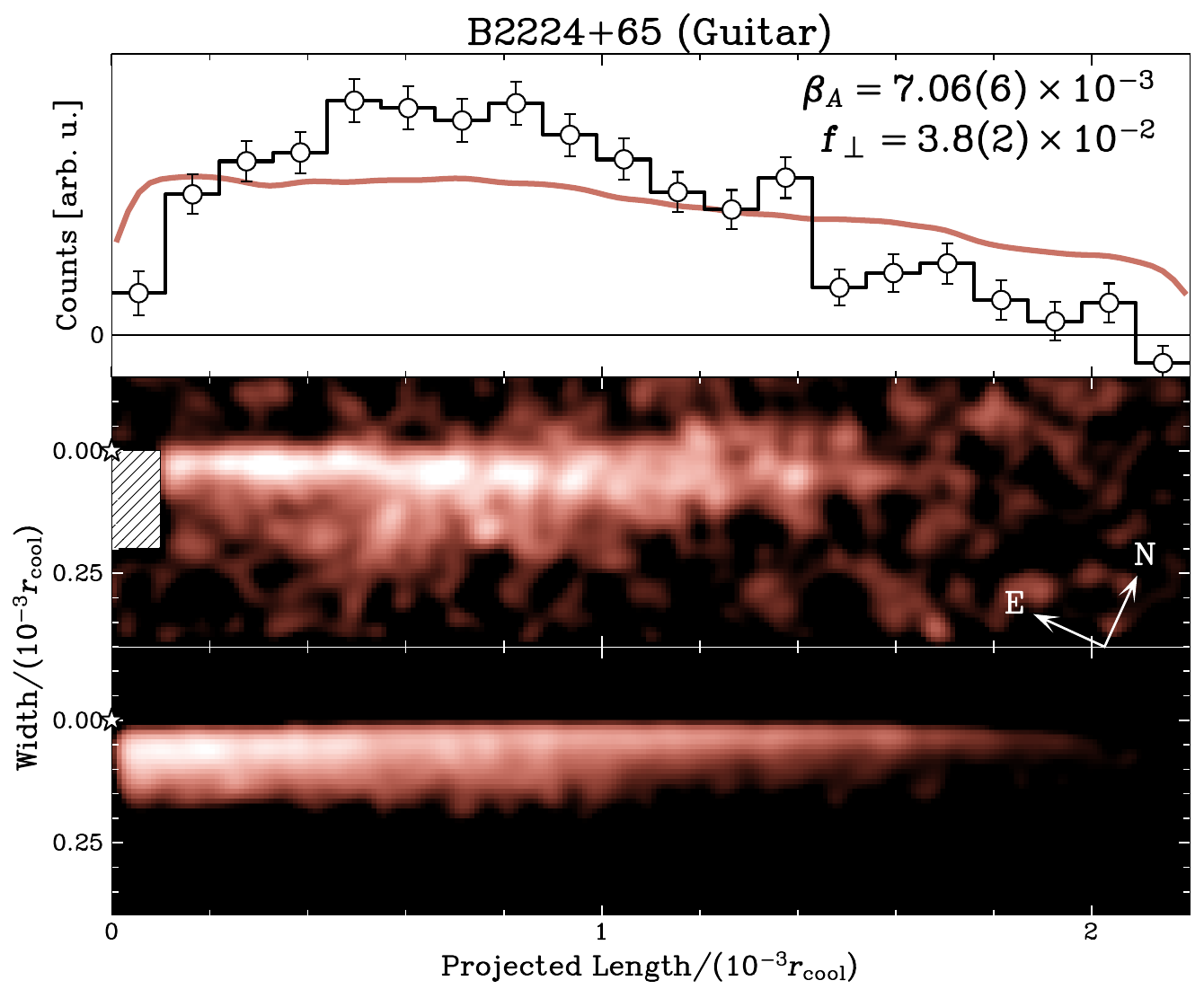}

  \includegraphics[width=0.95\linewidth]{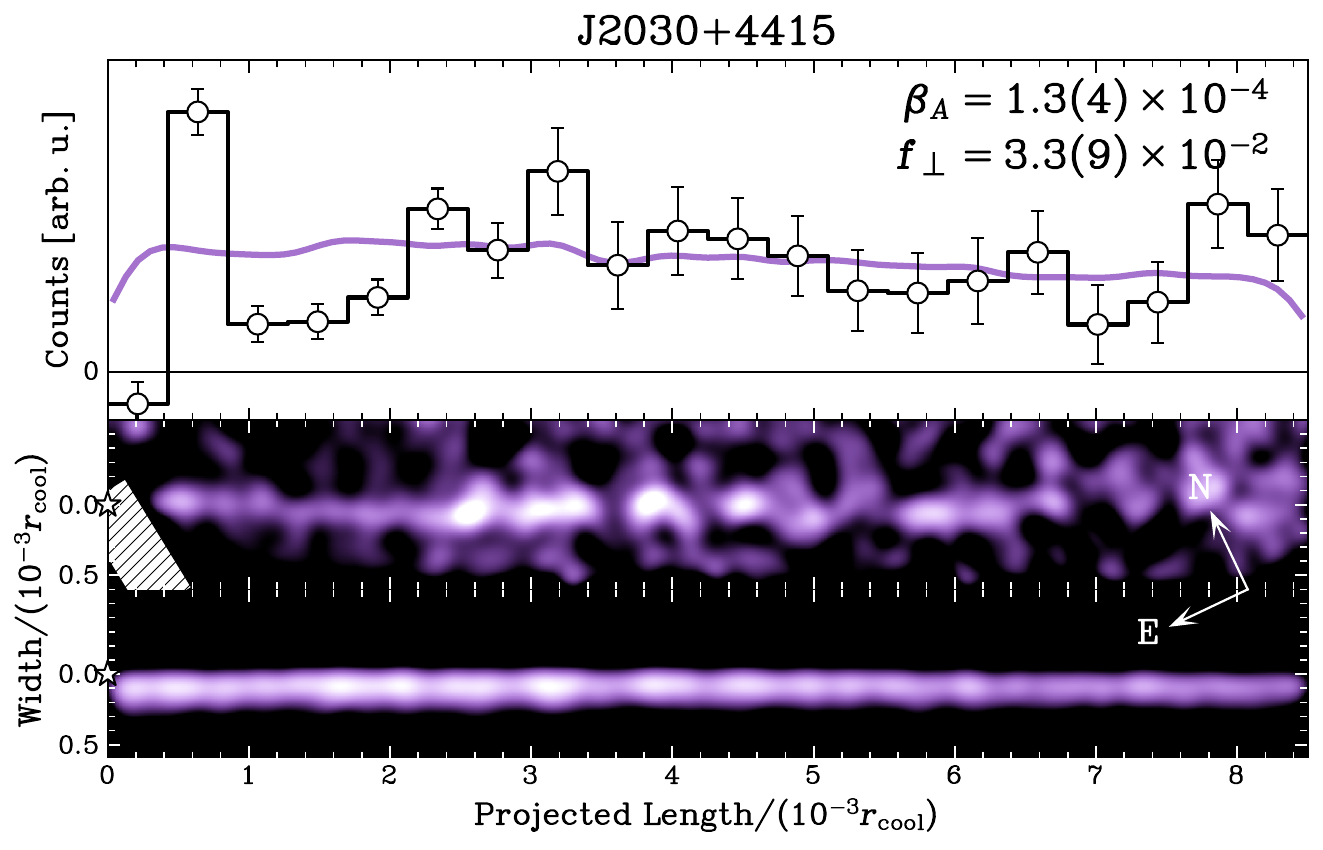}
  
  \includegraphics[width=0.95\linewidth]{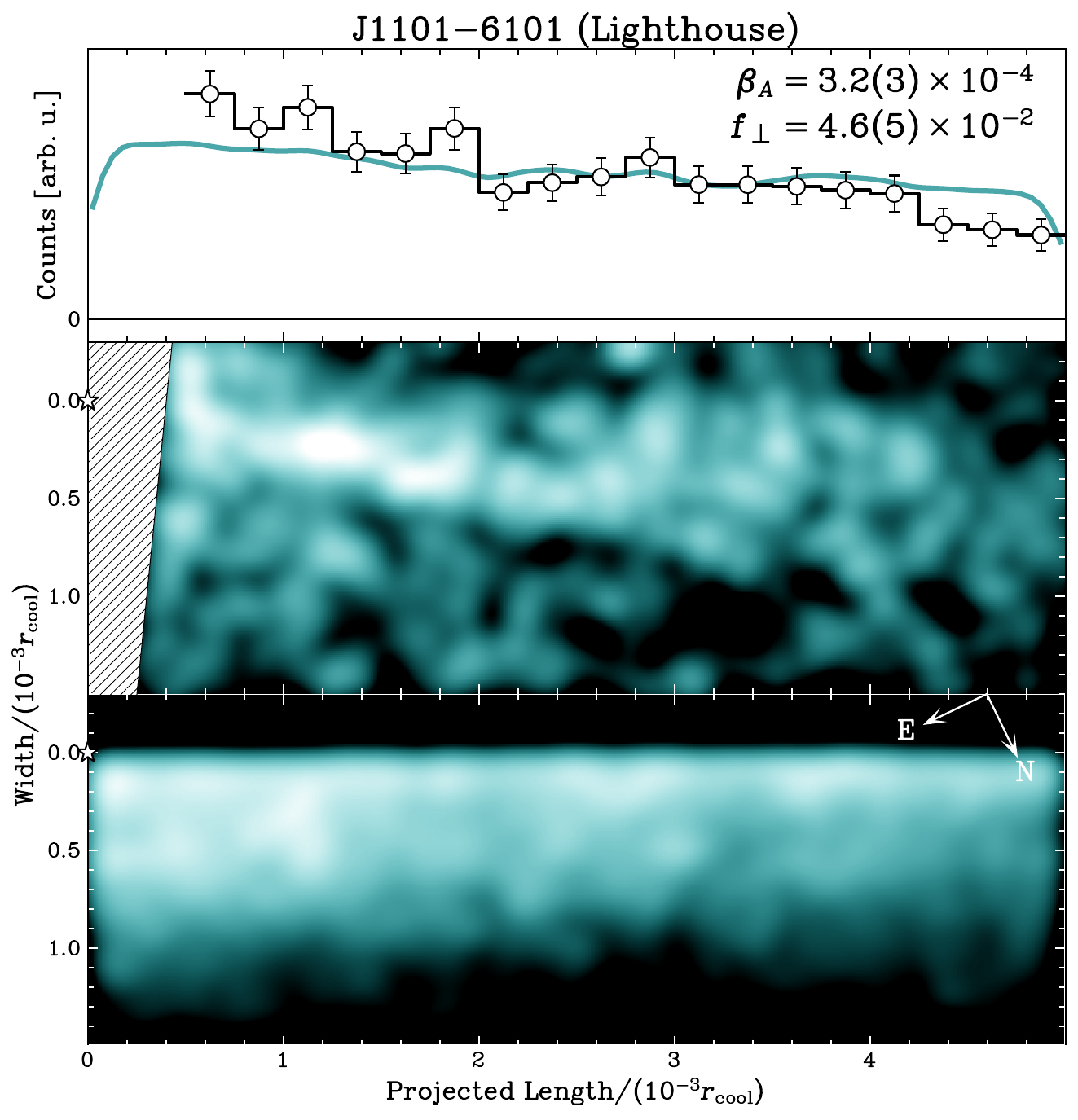}

  \caption{Comparisons between the model (bottom) and data (middle) for the three brightest X-ray filaments, assuming inclination $\iota=120^\circ$. Images and one-dimensional histograms (top) integrated over width are shown. The pulsar is moving up in all images.}
  \label{fig:data}
\end{figure}

\subsection{Choice of Parameters}
\label{sec:parameters}
Our model depends primarily on the injection rate $\dot E_\mathrm{inj}$, background field strength $B_0$, the ambient ISM diffusion constant $D_\parallel$, the Alfv\'en velocity $\beta_A$ (controlling the rate of turbulence generation), and the turbulence ratio between large and small scales $f_\perp$ (controlling perpendicular diffusion). However, $B_0$ is largely degenerate with the other parameters because it controls the characteristic length scales of $a_B$ and the cooling length. $f_\perp$ and $D_\parallel$ are also degenerate, as they both affect the rate at which reflecting particles pollute the injection zone. To avoid degeneracies, we fix $B_0$ and $D_\parallel$ to ISM values and fit $\beta_A$, $f_\perp$, and $\dot E_\mathrm{inj}$ to data.

Often quoted general ISM values are $B_0=3-5\ \mu$G, though filament equipartition values are closer to 15 $\mu$G \citep{dinsmore2024catalog}. Diffusion constants are observed in the range $D_\parallel = 1-5E_\mathrm{TeV}^{0.3-0.8}$ pc$^2$ yr$^{-1}$ \citep{dimitrakoudis2009obtaining, aharonian2007discovery, busching2007obtaining, castellina2005diffusion, cho2003compressible}. From the requirement of small bow shock standoff and the fact that at least three X-ray filaments are associated with H$\alpha$ bow shocks \citep[Guitar, J2030, and J1509$-$5850;][]{brownsberger2014survey}, we already know that filaments arise in the dense, colder WNM where the field is higher than that of the volume-dominating ionized ISM. The Galactic center presents a clear case of elevated magnetic field in a special region \citep[10 $\mu$G--1 mG,][]{ferriere2009interstellar}, and it contains a filament candidate. Therefore $B_0$ may be elevated above typical values, and if the fraction of energy in turbulence $u_\tau$ remains constant then $D_\parallel$ would be suppressed.\footnote{Eq.~\ref{eqn:d-cos-alpha} trends as $D_{\cos\alpha} \propto u_\tau B_0^{1/2}$ assuming $D_\parallel \propto E^{1/2}$ scaling. Since $D_\parallel \propto D_{\cos \alpha}^{-1}$ \citep{shalchi2005second}, $D_\parallel \sim u_\tau^{-1} B_0^{-1/2}$ so increasing $B_0$ decreases $D_\parallel$ at constant $u_\tau$.} We therefore fix $B_0 = 10$ $\mu$G and $D_\parallel = 1 E_\mathrm{TeV}^{0.5}$ pc$^2$ yr$^{-1}$.

Typical values of $B_0$ and $D_\parallel$ inferred for the (volume-dominating) ionized ISM reduce the turbulence generated in the leading edge and produce filaments that are too long. As described above, our selected values seem plausible for the denser neutral ISM; filaments found without associated H$\alpha$ bow shocks may tend to be longer. Alternatively our preference for high $B_0$ and low $D_\parallel$ may also stem from our assumption of linearity. If, via nonlinear processes, turbulent cascades transfer more efficiently from non-resonant scales to resonant scales or increase turbulence generation rates, this could relax the need for large $B_0$ or small $D_\parallel$.

\subsection{Fit to Data}
\label{sec:results}
We fit our filament model to \textit{Chandra} observations of Guitar, J2030, and Lighthouse---the best observed filaments and the only three with published transverse velocity measurements. \cite{dinsmore2024catalog} tabulates this velocity and the pulsar spin-down power. We assume zero radial velocity and set the field perpendicular to the pulsar velocity $\bm v_\mathrm{psr}$. The filament inclination angle to the line of sight $\iota$ affects the apparent filament morphology due to projection effects and beaming. The statistically averaged projection occurs at inclination angle $\iota \sim 60^\circ$ or $120^\circ$. We adopt the latter, since an initial CR stream beamed away from the line of sight tends to improve the fits.

The filaments are simulated for the cooling lifetime of a typical filament particle, ensuring steady state is reached. In the case of Lighthouse, we have fitted only the filament near the pulsar and ignored the larger downstream structure (see Fig.~\ref{fig:time}, bottom panel), which likely represents trapped particles from an earlier, more energetic injection epoch (\S \ref{sec:time}). We then compute the $\chi^2$ between observed filament images and the predicted synchrotron image, and use this to determine the best fit $\dot E_\mathrm{inj}$, $\beta_A$ and $f_\perp$. This approach ensures that the total synchrotron luminosity emitted by our filaments, computed as discussed near Eq.~\ref{eqn:sync}, matches observations. Details are given in appendix \ref{app:fit}.

Fig.~\ref{fig:data} shows a comparison between the best-fit simulated images and \textit{Chandra} data, both expressed as a two-dimensional image and a one-dimensional surface brightness profile integrated along the filament width. Units are 0.5--7 keV counts s$^{-1}$. The images have been rotated and flipped such that the pulsar is on the left edge and moves upwards. Fine details of the magnetic field structures etc.\,are not captured in this model, so the fit $\chi^2$ are $>1$. However, the fits reasonably constrain the parameters, with values displayed at the upper right in each panel.

Lighthouse and J2030 show reasonable agreement between the model and data. In particular, the correct length and width are reproduced, and the integrated histograms are matched. Far from the pulsar, the Lighthouse simulated image matches the data less well, though the simulation predicts $u_\tau$ approaching one in select zones behind the leading edge in this case. This calls our assumption of perturbative turbulence into question for Lighthouse. The filament front is straighter in all models than in the data, possibly due to (un-modeled) field line drift. The Guitar image matches the data well, though the integrated histogram predicts excess flux far from the pulsar. This excess could be addressed by adjusting the inclination angle (fixed here at the typical value) since projection effects and beamed radiation from CR are sensitive to this parameter.

\begin{figure}
  \centering
  \includegraphics[width=\linewidth]{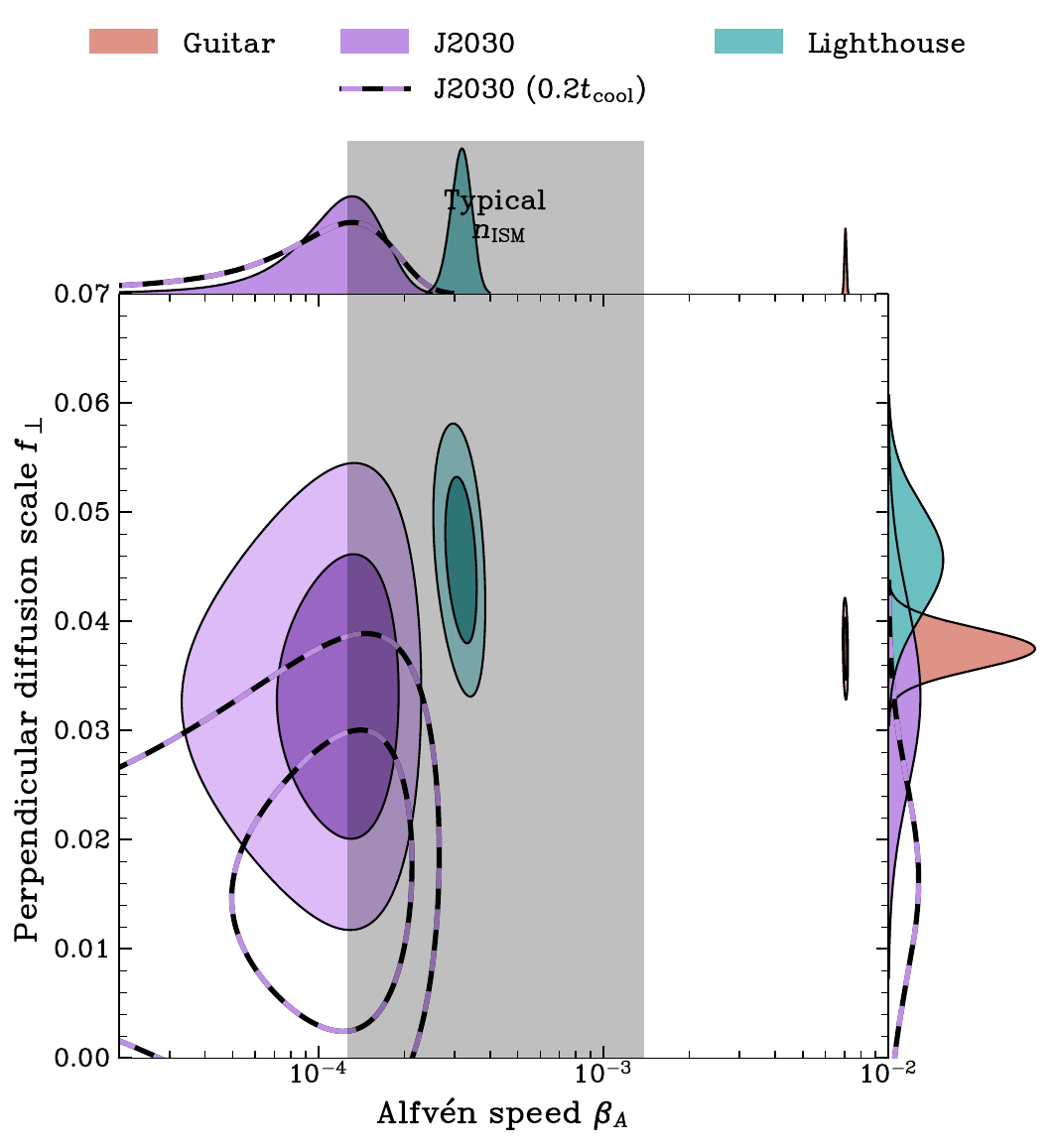}
  \caption{68 and 95\% posterior contours for our fit to observed images of Guitar, Lighthouse, and J2030. The band indicates typical ISM $\beta_A$. The Guitar may live in an especially low ionization zone (see text). All filaments appear to be well-described by similar $f_\perp$.}
  \label{fig:corner}
\end{figure}

Fig.~\ref{fig:corner} shows the Gaussian uncertainties associated with fit parameters $\beta_A$ and $f_\perp$ and compares them to typical ISM values. We compute typical Alfv\'en velocities assuming a WNM number density of $n_\mathrm{ISM} = 0.1-3$ cm$^{-3}$ and ionization fraction of 0.02--0.08 \citep{wolfire1995neutral, jenkins2013fractional}. The main tension is Guitar's large Alfv\'en velocity, which can be understood if the local ISM has an unusually low ionization \citep[as found in][]{brownsberger2014survey}, or if our $B_0=10\,\mu$G estimate is incorrect. All filaments are well-modeled with a uniform value of $f_\perp \approx 0.038$, indicating that the ratio of perpendicular to pitch-angle diffusion coefficient is fairly universal, insensitive to environment.  Altering the assumed inclination angle modestly changes the fit parameters.

Motivated by J2030's unusually narrow morphology, previous works have proposed that this filament is unusually young \citep{de2020psr,olmi2024nature}. We explore both a steady state scenario (solid contours of Fig.~\ref{fig:corner}) and a young filament with age $0.2t_\mathrm{cool}$ (hollow contours), where $t_\mathrm{cool}$ is the typical CR cooling time in $B_0$ fields. Both results are consistent with typical ISM $\beta_A$. The young model prefers a somewhat lower $f_\perp$ than that of other filaments.

\section{Predictions of the Model}
\label{sec:predictions}
\subsection{Filament Morphology and Time Dependence}
\label{sec:time}
\begin{figure}
  \centering
  \includegraphics[width=0.99\linewidth]{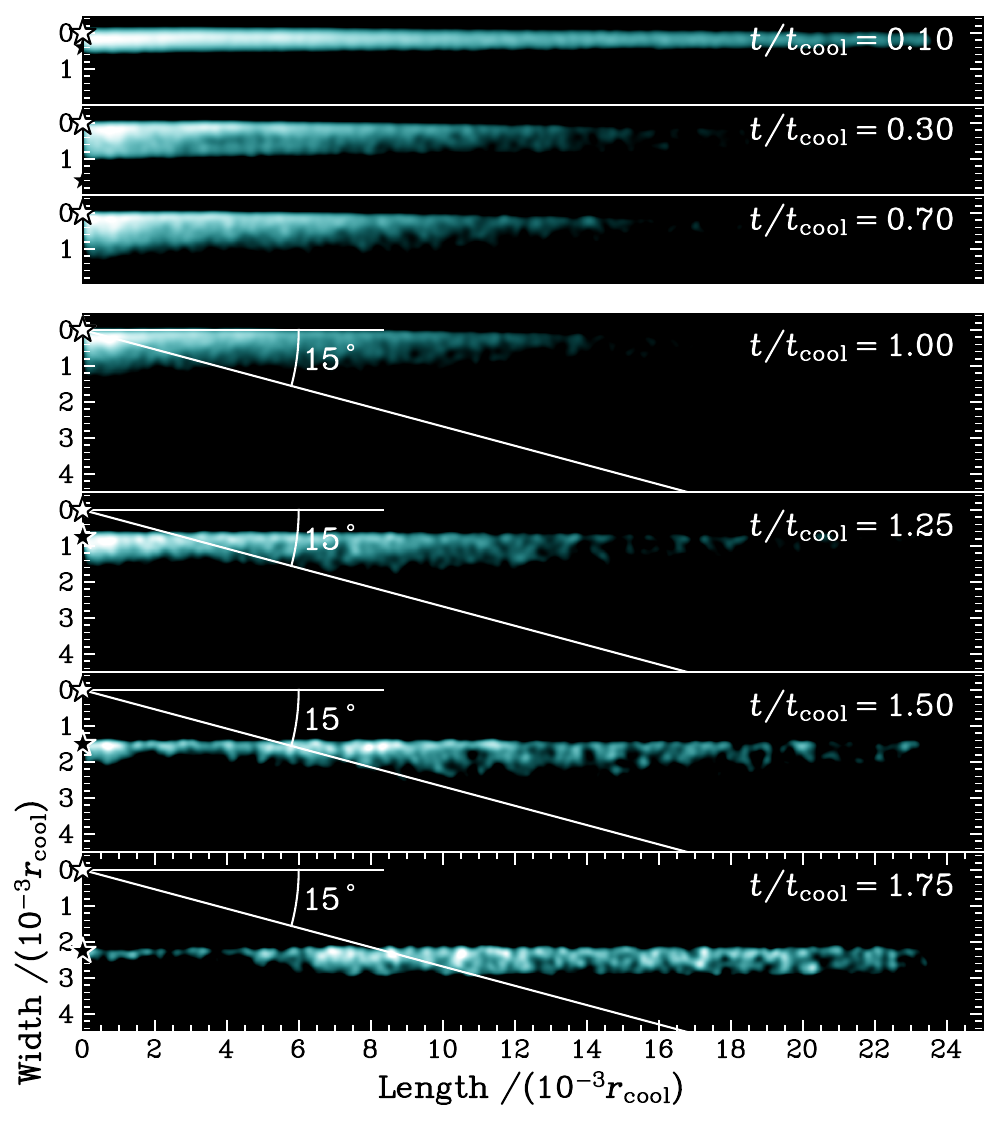}
  
  \vspace{0.5cm}
  \includegraphics[width=0.95\linewidth]{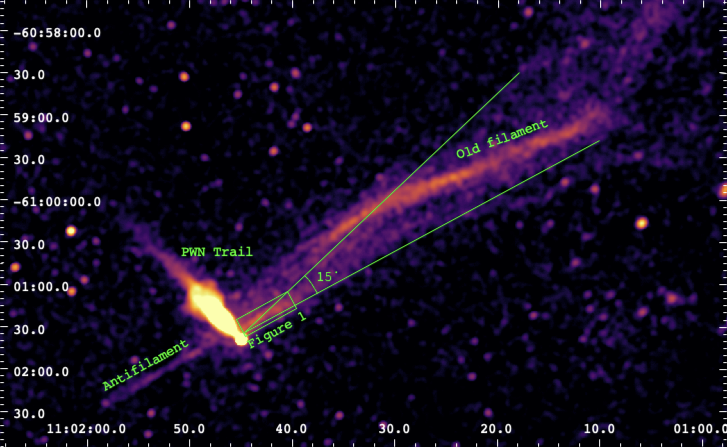}
  \caption{\textit{Top:} The Lighthouse model filament growth over time, with injection cutoff at $t=t_\mathrm{cool}$. Stars on the left edge indicate the pulsar's current position and position at cutoff. A reference line (15$^\circ$ from the pulsar position) shows how the peak of an old filament drifts outward with time. \textit{Bottom:} \textit{Chandra} data of the Lighthouse filament.}
  \label{fig:time}
\end{figure}

Our simulations predict filament three-dimensional structure and morphology variations over time, including aspects inaccessible to current observations. Approximate steady state is reached $\sim 0.3 t_\mathrm{cool} \approx 600$\,yr after injection is initiated, over which time the pulsar has traversed a large fraction of the filament width. The top three panels of Fig.~\ref{fig:time} show the Lighthouse model's development during this time. Variability of the particle injection rate over these 600 years could alter the brightness profile of the filament.

The simulations presented in \S\ref{sec:results} had steady energy injection, but previous work has suggested that filament energy injection can be variable \citep[e.g.][]{de2020psr}. We therefore explore a scenario where injection is short-lived. The next four panels of Fig.~\ref{fig:time} display the Lighthouse filament with injection cutoff at $t/t_\mathrm{cool}=1$. As particles cool, the filament remains anchored to its original ISM field lines, lagging behind the pulsar and drifting away from the pulsar's trail. The angle between the net drift velocity and the leading edge is $\sim 15^\circ$ in the filament-trail plane. This drift could explain the large structure labeled ``old filament'' in the bottom panel of Fig.~\ref{fig:time} which shows \textit{Chandra} observations of the Lighthouse. The simulated luminosity of the cooling remnant halves every $0.25 t_\mathrm{cool}$ for these parameters, reaching 10\% of its pre-cutoff flux in the last panel. To reproduce the large flux of the observed ``old filament,'' the previous injection period must have exhibited a much larger particle flux than the current period, or un-modeled effects such as magnetic mirroring from ISM field strength variations suppress particle escape in this older region.

\subsection{Filament Spectrum and Particle Escape}
\label{sec:escape}
As mentioned in step E of \S\ref{sec:argument}, a significant fraction of CRs escape from the filament's far end during their random walk in the turbulent fields behind the leading edge. This escape is energy dependent, and therefore also affects the observed photon spectrum. Our simulation predicts the spectrum and escape fraction purely from the methods introduced in \S\ref{sec:model}. However, we can also find these quantities using an analytical spectrum model with one free parameter (other than normalization) which we now introduce. An analytical model for spectrum and escape fraction is possible because most of the filament luminosity comes from CRs behind the leading edge where turbulence is well developed and smooth. We suppose that, in this smooth turbulence region, the probability to escape via random walk $P_\mathrm{esc}$ is constant in time. One expects $P_\mathrm{esc}$ to depend on CR energy through the energy dependence of $D_{\cos \alpha}$. We estimate $P_\mathrm{esc}(\gamma) \propto \gamma$ by monitoring the energy-dependent escape rate of particles in our simulation of Guitar, and indeed the following results calculated assuming $P_\mathrm{esc}(\gamma) \propto \gamma$ will be shown to be consistent with our simulation.

The steady-state particle spectrum as a function of Lorentz factor $dN/d\gamma=f(\gamma)$ is shaped by $P_\mathrm{esc}(\gamma)$, the particle injection spectrum $Q_\mathrm{inj}(\gamma)$, and radiative synchrotron loss transferring CRs to lower energies at a rate $\propto \gamma^2$. Conservation of particle number yields the differential equation
\begin{equation}
    -\frac{\partial}{\partial\gamma} \brackets{f(\gamma) \gamma^{2}} \propto Q_\mathrm{inj}(\gamma) - P_\mathrm{esc}(\gamma) f(\gamma).
    \label{eqn:spectrum-diffeq}
\end{equation}
Our injection spectrum (\S\ref{sec:injection}) is $Q_\mathrm{inj}(\gamma)\propto \gamma^{-p}$ above a minimum Lorentz factor $\gamma_\mathrm{gate}$ with $Q_\mathrm{inj}(\gamma < \gamma_\mathrm{gate}) = 0$. Given this, the solution to Eq.~\ref{eqn:spectrum-diffeq} is
\begin{equation}
  f(\gamma) \propto \begin{cases}
    \gamma^{-s}, & \mathrm{for}\ \gamma \leq \gamma_\mathrm{gate}\\
    \gamma^{-(p+1)}  & \mathrm{for}\ \gamma \geq \gamma_\mathrm{gate}
  \end{cases}.
  \label{eqn:particle-spec}
\end{equation}
The index $s$ is related to the normalization of $P_\mathrm{esc}$; it is the model's free parameter. For $s<0$, this hard low-energy power law  ensures that the spectrum does not extend far below the soft X-rays. Above $\gamma_\mathrm{gate}$, the spectrum appears entirely ``cooled'' with no cooling break.

The fraction of injected energy that is carried away by escaping particles $f_\mathrm{esc}$ is equal to the total escape rate divided by the injection rate, $f_\mathrm{esc} = \int d\gamma\, P_\mathrm{esc}(\gamma) f(\gamma) / \int d\gamma\, Q_\mathrm{inj}(\gamma)$. This integral depends on the normalization of $P_\mathrm{esc}(\gamma)$, which depends on $s$. Using Eq.~\ref{eqn:particle-spec}, $f_\mathrm{esc}$ can be shown to take the form.
\begin{equation}
    f_\mathrm{esc} = \frac{2-s}{\mathrm{min}(3, 1+p)-s}.
    \label{eqn:escape-frac}
\end{equation}

\begin{figure}
  \centering
  \includegraphics[width=\linewidth]{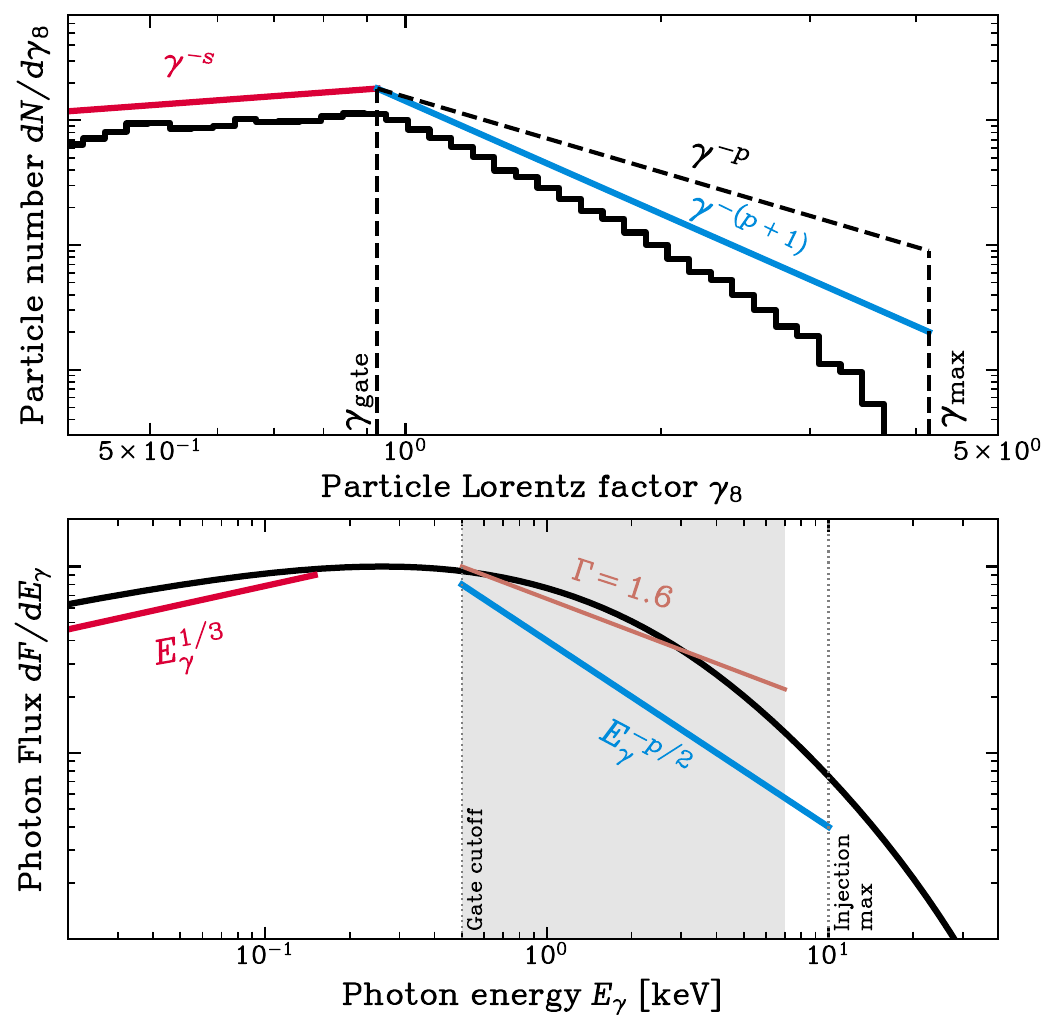}
  \caption{\textit{Top:} The histogram shows the steady state particle spectrum for the Guitar filament simulation. The dashed line indicates the injected spectrum, and solid colored lines are reference power laws described in the text. \textit{Bottom:} The synchrotron photon spectrum, with reference power law components. The photon index $\Gamma$ for a power law fit to the segment of the spectrum in the \textit{Chandra} band (shaded) is consistent with measurements (orange line).}
  \label{fig:spectrum}
\end{figure}

In the top panel of Fig.~\ref{fig:spectrum}, we show the particle spectrum observed in our simulation. It shows good agreement with the model spectrum $f(\gamma)$ for $s = -1/2$. With this $s$ and our injected $p=2$, Eq.~\ref{eqn:escape-frac} gives  $f_\mathrm{esc} = 71\%$. We independently measure $f_\mathrm{esc}$ from the Guitar simulation by tracking the rate of CRs that exit the simulation box and dividing by the simulation injection rate $\dot E_\mathrm{inj}$. The result is $f_\mathrm{esc}=69\%$ --- a valuable numerical confirmation of our simple spectrum model. 

\label{sec:spectrum}

The remaining fraction $1-f_\mathrm{esc}$ cools via radiation and gives the filament luminosity $L_\mathrm{fil}$. The rate of energy injected by the pulsar is $\dot E_\mathrm{inj}= L_\mathrm{fil} / (1-f_\mathrm{esc})$, and the unseen power carried away by escaping particles is $\dot E_\mathrm{esc} = L_\mathrm{fil} f_\mathrm{esc}/(1-f_\mathrm{esc})$. Note that observed filament luminosities are far below the pulsar spin down luminosity $\dot E$: $L_\mathrm{fil} \sim 10^{-4}-10^{-3} {\dot E}$ \citep[e.g.][]{dinsmore2024catalog}. The escaping / injected power predicted by this simplified spectral model is therefore far below $\dot E$---within the pulsar energy budget---unless $f_\mathrm{esc} > 99.9\%$.

The bottom panel of Fig.~\ref{fig:spectrum} reports the simulated synchrotron spectrum, smoothed by synchrotron emissivity and the pitch angle distribution. Thus, if $\gamma_\mathrm{gate}$ is as high as we have suggested, the photon spectrum is not a strict power law. The measured $\Gamma$ (defined by $dN_\gamma/dE_\gamma \propto E_\gamma^{-\Gamma}$) represents an approximation over the {\it Chandra} band. For our parameter choices, the overall photon index agrees with observed values.

This analysis suggests a potentially powerful technique to understand filament processes. If $s$ and $p$ are accurately measured from the X-ray spectrum, then the escape and injection rates $\dot E_\mathrm{esc}$ and $\dot E_\mathrm{inj}$ can be estimated. Since the unabsorbed X-ray band accessible to \textit{Chandra} is quite limited, no significant deviations from power law spectra have yet been measured. If more observations thoroughly exclude a spectral break, then $\gamma_\mathrm{gate}$ must be low so that the $dF/dE_\gamma = E^{-p/2}$ power law dominates the \textit{Chandra} band (i.e.~$\Gamma = p/2+1$). A very hard $p=1$ injection spectrum would then be necessary to match observed photon indices of $\Gamma \approx 1.5$. This would be useful in constraining the processes that accelerate the filament CRs. Radio and optical upper limits on the spectrum flux are useful, but we do not expect these to probe the details of the injected spectrum as the $E_\gamma^{1/3}$ synchrotron emission from $\gamma_\mathrm{gate}$ particles dominates at the lowest energies. Very sensitive observations could also be useful in detecting the faint, hard-spectrum emission expected from particles escaping past the end of the filament.

\begin{figure}
  \centering
  \includegraphics[width=\linewidth]{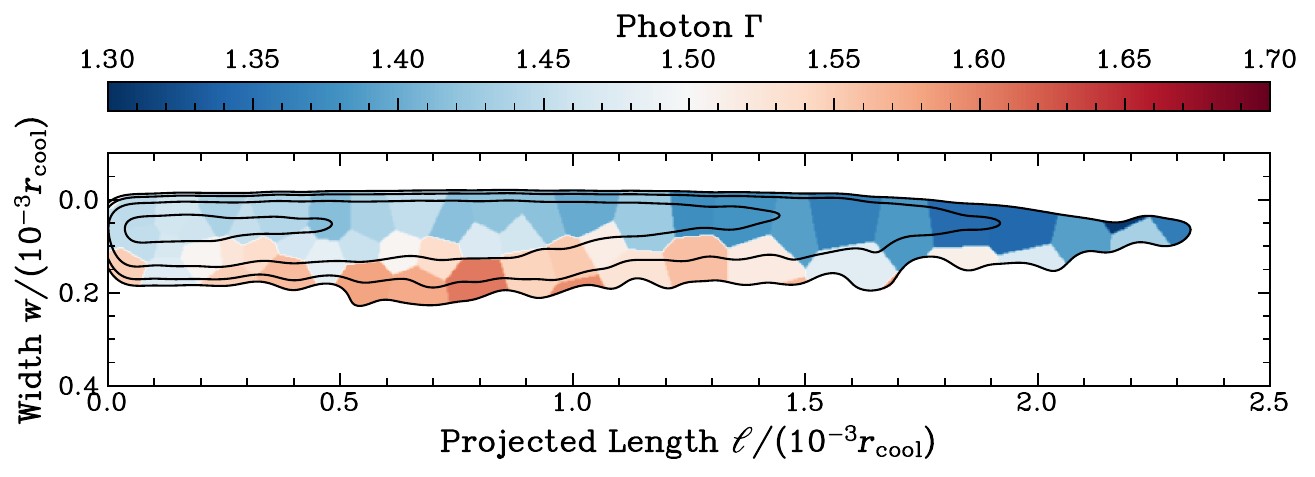}
  \caption{Photon spectral indices $\Gamma$ fitted to the Guitar filament model as a function of position.}
  \label{fig:voronoi}
\end{figure}

Fig.~\ref{fig:voronoi} shows the spatial dependence of the photon index over the simulated Guitar filament. The index is obtained by generating the synchrotron spectrum from the particles in each bin through Eq.~\ref{eqn:sync-spec} and fitting a power law. Particles behind the filament are older and softer while the front is quite hard. This effect has been observed in Guitar \citep{de2022quarter}. Our model predicts little spectral change along the filament length. Observations are consistent with this, though they prefer slight softening at large distance \citep{de2022quarter}. Cooling behind the filament edge is difficult to measure, but is expected in our picture and may be accessible to future X-ray observatories.

\subsection{TeV Halos and the Positron Excess}
\label{sec:collapse}
TeV/GeV observations have revealed ``halos'' around Geminga \citep{abdo2009milagro}, PSR B0656+14 \citep[Monogem,][]{2017abeysekara2hwc}, and several other energetic  $\dot E > 10^{36}$ erg s$^{-1}$ pulsars \citep{2023PhRvD.107l3020S}. They are now expected to exist around most middle-aged, energetic pulsars \citep{2025PhRvL.134q1005A}. These spherical, extended sources of Compton emission from $\sim$ TeV particles imply a diffusion coefficient $\sim 100\times$ smaller than the typical ISM value \citep{linden2017using}, rendering particle escape from these sources inefficient. CR-generated turbulence may help explain this decreased diffusion \citep{evoli2018self,mukhopadhyay2022self}.

Filament collapse is expected for the energetic, low velocity TeV halo pulsars.  
Applying our filament model, their higher $\dot E$ and consequently higher CR injection rate increases the rate of turbulence generated by reflecting CRs (step B). Their lower velocity slows motion to new field lines, allowing continued turbulence growth (step C). The post-cutoff leading-edge turbulence will therefore be much larger, reaching the non-linear regime. Perpendicular diffusion then becomes rapid, creating a turbulent, isotropic PWN that efficiently traps particles instead of a leaky, filamentary structure. This isotropic PWN would contribute to an older, halo-emitting population confined near the pulsar. Our simulation cannot fully model the halos because non-linear turbulence violates assumption IV, but when we boost the turbulence level to approach non-linearity, we indeed find that nearly all particles are isotripised in a small, turbulent zone near the pulsar. High-velocity pulsars avoid this collapse by rapidly shifting the injection zone to clean magnetic field lines, so the post-cutoff turbulence is still linear \cite[e.g.~the Lighthouse pulsar, where $\dot E = 1.4 \times 10^{36}$ erg s$^{-1}$, but $v_\mathrm{psr} \geq 990$ km s$^{-1}$][]{dinsmore2026chandra}. We therefore predict that filament pulsars will not show spherical halos, but instead should show faint TeV emission trailing the pulsar filament.

The 300\,GeV positron excess discovered at Earth by PAMELA \citep{adriani2009anomalous}, \textit{Fermi} \citep{ackermann2012measurement}, and AMS \citep{aguilar2013first} has been suggested as a possible signal of dark matter annihilation \citep{cholis2009high}. More prosaically, the positrons could be pulsar-generated \citep{yuksel2009tev}, if such particles can reach Earth before they over-cool. Several studies have modeled positrons from known pulsars \citep[e.g.][and references therein]{2025JCAP...02..029O}. The situation is complicated by TeV halos since reduced diffusion lowers the total distance traveled before the positrons cool. While two-zone models appear to help match the observed excess spectrum above $\sim 1$\,TeV \citep{2025arXiv250317442J}, they require higher pulsar positron injection efficiency to match the observed local flux.

Under our weak turbulence filament model, filament-producing pulsars are unique in allowing a significant fraction of their injected high energy $e^\pm$ to escape to ISM magnetic field lines ($f_\mathrm{esc} \sim 70\%$, \S\ref{sec:escape}). There, a $\sim \gamma_8$ particle diffusing with  $D_\parallel = 1-5$ pc$^2$ yr$^{-1}$ travels $1-2B_{10}^{-2}$ kpc before cooling past the 300\,GeV excess peak \citep{aguilar2019towards}. Several filament pulsars (Guitar, J2030, PSR J2055+2539, and PSR J1957+5033) lie inside this horizon. Thus, while filament pulsars represent a small fraction of the total injected pulsar $e^\pm$ power, they may make an out-sized contribution to the positrons that reach Earth, especially at high energies. Given the slow rate of diffusion perpendicular to ISM field lines, the Earth positron flux could be disproportionately sensitive to the few pulsars connected to Earth by the local kpc-scale magnetic field lines. Improved mapping of such fields should allow us to follow positron propagation from nearby pulsars and evaluate the filament contribution to the local positron flux.

\section{Conclusions}
\label{sec:conclusion}

We have presented a new model of pulsar X-ray filaments in which CRs are partially trapped by perturbative, self-generated turbulence. This turbulence is mostly generated in leading edge of the filament where CRs are injected. Turbulence growth is initially due to the NRSI, whose small scale power weakly scatters CRs. However, a small amount of CR reflection dilutes the current and leads to a feedback loop creating resonant-scale RSI turbulence, which causes more CR reflection. This feedback loop exponentially grows turbulence until it is cut off (while still weak) as the pulsar moves to new field lines. The reflected CRs then diffuse in this post-cutoff turbulence, synchrotron radiating and eventually escaping into the ISM. A semi-analytical simulation built on these principles yields images similar to the Guitar, Lighthouse, and PSR J2030+4415 filaments and reproduces their observed spectra. Their best-fit physical ISM parameters are consistent with measured values except for a large Alfv\'en speed near Guitar, which we discuss. Synchrotron beaming is important and controls whether the narrow outflow of young particles is visible. Guitar in particular is better modeled under the assumption that the filament points away from the observer. This would beam the anti-filament's radiation toward Earth, and may explain its relatively large anti-filament-to-filament flux ratio.

Of course this model is schematic, and non-uniform background magnetic fields, second-order turbulence effects, etc.\,will modify the appearance of individual filaments.  Improved modeling of these effects can make filaments an interesting probe of ISM properties and allow extraction of their three-dimensional orientation. To improve the fits, we assume a slow inverse cascade to large scale turbulence, and we suppose that the injected particles are partly charge-separated to initiate NRSI turbulence. Higher fidelity simulations can probe whether these assumptions are required. For example, recent simulations by \citep{2025arXiv251215847O}, indicate that charge separation and NRSI turbulence growth can occur even for symmetric $e^\pm$ CR injected to a baryon-dominated ISM.

Our model exhibits several new properties. If injection is interrupted, an old cooling filament drifts away from the pulsar, producing structure reminiscent of Lighthouse's complex morphology. For filament pulsars the steady state turbulence is weak enough that approximately 70\% of the injected energy escapes along unperturbed ISM field lines. With the very hard injected particle spectra, this produces additional spectral features in the filament emission. Future high sensitivity filament X-ray spectra can thus probe the particle acceleration physics and measure the escape efficiency. Finally we note that this escape partly circumvents the scattering-induced CR confinement inferred from the GeV/TeV halos around nearby energetic pulsars; the filament pulsars may thus make surprisingly large contributions to the local positron excess.

\begin{acknowledgments}
The authors are grateful to Alexander A. Philippov for a careful reading and helpful comments. This work was funded in part by NASA grant G03-2404X administered by the Smithsonian Astrophysical Observatory.
\end{acknowledgments}

\vspace{5mm}
\facilities{\textit{Chandra} X-ray Observatory (CXO)}
\software{\texttt{Rust}, \texttt{Python}, \texttt{ciao} \citep{fruscione2006ciao}}

\bibliography{bib}{}
\bibliographystyle{aasjournal}

\appendix

\section{Turbulent magnetic field geometry}
\label{app:b}
This appendix defines the turbulent spectral energy density $s_\tau(k)$ and derives the boost it causes to the synchrotron luminosity. Writing $\bm b(\bm k)$ as the Fourier transform of $\bm b(\bm x)$,\footnote{Our Fourier transform convention is $\bm b(\bm k) = \int d^3 \bm x\, e^{-i\bm k \cdot \bm x} \bm b(\bm x)$.}
the geometry is summarized by the correlation functions of $\bm b(\bm k)$. In Fourier space, $\nabla \cdot \bm B = 0$ is written as $\bm b \cdot \bm k = 0$. Our slab geometry therefore requires that $\bm k = k \unit z$, so the correlation functions only depend on a scalar variable $k$.

We further assume straight magnetic field lines so that the average value of $\bm b$ is $\chevronsi{\bm b(k)} = 0$, and parity and rotational symmetry on average so that $\chevronsi{b_i(k) b^*_j(k')} = 0$ when $i\neq j$. The remaining correlation function is $\chevronsi{b_i(k) b^*_i(k')}$ for $i=1$ or $2$, which we assume are equal. We express this function using the turbulence energy density $s_\tau(k)$:
\begin{equation}
  \chevronsi{b_i(k) b^*_j(k')} = \pi^2\delta_{ij} \delta(k - k') s_\tau(k). \qquad i = 1,2
  \label{eqn:tensor}
\end{equation}
The $\pi^2$ normalization is inserted so that Eq.~\ref{eqn:u_tau} holds.

A relativistic particle with velocity $\beta$ in units of $c$ orbiting in this magnetic field accelerates as \begin{equation}
    \dot {\bm \beta} = \omega_B \bm \beta \times (\unit z + \bm b).
    \label{eqn:beta-accel}
\end{equation}
The Larmor dipole formula predicts that this acceleration leads to radiative power $L = 2e^2 \gamma^4 \dot \beta^2/(3c)$ \citep{rybicki2024radiative}. We have implicitly used the fact that the acceleration is perpendicular to $\bm \beta$ because static magnetic fields do no work. Applying Eq.~\ref{eqn:beta-accel} to this formula, the square of the $\unit z$ term gives the standard synchrotron luminosity for the background field and the square of the $\bm b$ term gives the turbulent correction. The cross term does not contribute because $\chevronsi{\bm b} = 0$.

To compute the turbulent correction, we note that the acceleration caused by turbulence is $\omega_B^2 \chevronsi{(\bm \beta \times \bm b)^2} = \omega_B^2 (\chevronsi{b^2} - \chevronsi{(\bm \beta \cdot \bm b)^2})$ by a vector identity. The first term emits power in proportion to the total magnetic energy $\chevronsi{\bm b^2} = u_\tau$. The second accounts for alignment between the velocity and the turbulent field. Since $\bm b$ is transverse with energy distributed equally between $\unit x$ and $\unit y$, we have $\chevronsi{(\bm \beta \cdot\bm b)^2} = u_\tau \sin^2 \alpha / 2$. Inserting these values into the dipole formula gives Eq.~\ref{eqn:sync} in the main text.

\begin{table*}
    \centering
    \begin{tabular}{lcc|lcc|lcc}
        \hline \hline
        Obs \# & Date & Exp. [ks]   & Obs \# & Date & Exp. [ks]  & Obs \# & Date & Exp. [ks] \\ \hline 
        &\textit{PSR B2224+65}&     &24431 & 2021-04-23 & 26     &22172 & 2019-04-14 & 44\\
        755   & 2000-10-21 & 49     &24432 & 2021-11-14 & 30     &22173 & 2019-04-15 & 22 \\ 
        6691  & 2006-08-29 & 10     &24433 & 2021-04-21 & 26     &23536 & 2021-02-13 & 25 \\ 
        7400  & 2006-10-06 & 37     &24434 & 2021-03-15 & 30     &24236 & 2021-11-08 & 29\\ 
        13771 & 2012-08-01 & 49     &24435 & 2021-03-15 & 15     &24954 & 2021-02-13 & 20 \\ 
        14353 & 2012-07-29 & 35     &24436 & 2021-04-03 & 25     &&\textit{PSR J1101-6101}&\\ 
        14467 & 2012-07-28 & 15     &24437 & 2021-02-19 & 25     &12420 & 2011-09-06 & 5 \\ 
        23537 & 2021-10-20 & 57     &24992 & 2021-03-16 & 15     &13787 & 2012-10-11 & 49 \\ 
        24426 & 2022-02-21 & 21     &26336 & 2022-02-24 & 18     &16007 & 2014-08-28 & 116 \\ 
        24427 & 2021-07-25 & 25     &&\textit{PSR J2030+4415}&   &16517 & 2014-09-05 & 52 \\ 
        24428 & 2021-07-04 & 30     &14827 & 2014-04-15 & 25     &16518 & 2014-09-29 & 10 \\ 
        24429 & 2021-04-25 & 25     &20298 & 2019-04-10 & 44     &17421 & 2014-10-02 & 20\\ 
        24430 & 2021-10-09 & 30     &22171 & 2019-04-12 & 40     &17422 & 2014-10-01 & 49 \\ 
        \hline \hline
    \end{tabular}
    \caption{Details of the filament observations used in this work. Observation numbers, dates, and exposures are listed.}
    \label{tab:obs}
\end{table*}

\section{Fit to Data}
\label{app:fit}

After reprocessing the filament \textit{Chandra} observations listed in Table \ref{tab:obs} with the standard \texttt{ciao} version 4.16 tools \citep{fruscione2006ciao}, point sources are carefully subtracted using the method of \cite{dinsmore2024catalog} and observations are realigned to track the pulsar's proper motion. After masking the pulsar and its PWN, the 0.5--7 keV data are spatially binned in a wide region around the filament and assigned $\sqrt{N}$ errors. Slight, chip-by-chip corrections to the \texttt{ciao} exposure maps are necessary to make the background uniform. We then exposure-correct the observations and subtract background.

Simulated observations are generated by binning the filament particles with the same grid. Predicted count maps are computed by calculating the number flux in the observable band using Eq.~\ref{eqn:sync-spec}, and multiplying by the effective area and quantum efficiency of the deepest observation. Goodness of fit is evaluated by the $\chi^2$ statistic. In this fit, $\beta_A$ and $f_\perp$ are adjusted to match observed filament morphologies by minimizing $\chi^2$. The simulated filament luminosity is then measured and compared to the observed luminosity, and $\dot E_\mathrm{inj}$ is correspondingly adjusted. The process is repeated until convergence. This fit method guarantees that the best-fit filaments produce luminosities that agree with observations.

As a diffusion simulation, our model predictions exhibit randomness which is important near the minimum $\chi^2$ model. We therefore evaluate the model on a grid near the minimum $\chi^2$ and fit a paraboloid to estimate Gaussian uncertainties. The residuals reveal a imperfect fit, with $\chi^2$ per degree of freedom of 1.3, 3.9, and 2.2 for Guitar, J2030, and Lighthouse. This is to be expected, since the model does not include many effects that must alter the detailed filament morphology, including wandering background field lines, terms second-order in $u_\tau$, and variations in $f_\perp$, $\beta_A$, and the injection strength over the filament lifetime. As an approximation to the systematic uncertainties incurred by leaving out these effects, we rescale uncertainties such that $\chi^2/\textrm{dof} = 1$.

\end{document}